\documentclass[12pt]{article}

\newcommand{\be}{\begin{eqnarray}}
\newcommand{\ee}{\end{eqnarray}}
\newcommand{\non}{\nonumber}
\newcommand{\tr}{\mathop{\rm tr}\nolimits}
\newcommand{\n}{\ensuremath{\mathcal{N}}}
\newcommand{\R}{\ensuremath{\mathsf{R}}}
\newcommand{\K}{\ensuremath{\mathsf{K}}}
\newcommand{\M}{\ensuremath{\mathsf{M}}}
\newcommand{\aaa}{\ensuremath{\mathsf{a}}}
\newcommand{\bbb}{\ensuremath{\mathsf{b}}}
\newcommand{\ccc}{\ensuremath{\mathsf{c}}}

\begin{document}

\begin{titlepage}
\strut\hfill UMTG--219
\vspace{.5in}
\begin{center}

\LARGE Soliton $S$ matrices for the critical $A_{\n-1}^{(1)}$ 
chain\\[1.0in]
\large Anastasia Doikou and Rafael I. Nepomechie\\[0.8in]
\large Physics Department, P.O. Box 248046, University of Miami\\[0.2in]  
\large Coral Gables, FL 33124 USA\\

\end{center}

\vspace{.5in}

\begin{abstract}
We compute by Bethe Ansatz both bulk and boundary hole scattering 
matrices for the critical $A_{\n-1}^{(1)}$ quantum spin chain.  The 
bulk $S$ matrix coincides with the soliton $S$ matrix for the 
$A_{\n-1}^{(1)}$ Toda field theory with imaginary coupling.  We verify 
our result for the boundary $S$ matrix using a generalization of the 
Ghoshal-Zamolodchikov boundary crossing relation.
\end{abstract}

\end{titlepage}

\section{Introduction}

Magnetic chains associated with affine Lie algebras \cite{bazhanov}, 
\cite{jimbo} constitute large classes of integrable models.  These 
models hold both mathematical and physical interest.  Indeed, the 
models can be shown to have certain quantum-algebra symmetries which 
are characterized by a parameter $q$.  A variety of powerful 
analytical techniques can be brought to bear on these models, yielding 
in the thermodynamic limit exact properties such as excitation 
spectra, scattering matrices, and correlation functions.  These 
models, which can be regarded as discretized integrable quantum field 
theories, have numerous applications to statistical mechanics and 
condensed matter physics.

In a previous paper \cite{DN0} we began investigating the simplest 
such class of models, namely, the $A_{\n-1}^{(1)}$ spin chain.  This 
class includes the well-known XXZ chain as the special case $\n=2$.  
There we considered the noncritical regime (real $q$), for which there 
is a mass gap, and for which the $S$ matrices are characteristic of 
$q$-deformed quantum field theory.

In the present work we continue our investigation of the 
$A_{\n-1}^{(1)}$ spin chain, but consider instead the critical regime 
($|q|=1$).  We compute by Bethe Ansatz both bulk and boundary hole 
scattering matrices.  Our result for the bulk $S$ matrix coincides 
with the soliton $S$ matrix for the $A_{\n-1}^{(1)}$ Toda field theory 
with imaginary coupling \cite{BL}-\cite{gandenberger1}.  We also 
generalize the Ghoshal-Zamolodchikov \cite{GZ} boundary crossing 
relation to the $A_{\n-1}^{(1)}$ case (for which the bulk $S$ matrix 
does not have crossing symmetry), and use it to help verify our result 
for the boundary $S$ matrix.  For $\n=2$, the boundary $S$ matrix 
reduces to the well-known result for the critical XXZ open spin chain 
\cite{GZ}-\cite{DN1}; while in the isotropic limit, it 
coincides with the result found in \cite{DN2}.

\section{The model and its symmetries}

We begin by briefly reviewing the construction of the model, both 
closed (periodic boundary conditions) and open with diagonal boundary 
fields.  \footnote{Our notations closely follow those of Ref.  
\cite{DN0}, in which the noncritical regime is studied, with the 
anisotropy parameter $\eta$ replaced by $i \mu$.} For an introduction 
to the Yang-Baxter equation, integrable spin chains, and the Bethe 
Ansatz, see e.g.  \cite{ABA}, \cite{devega}.

\subsection{Closed chain}

The main ingredient for constructing an integrable closed spin chain 
is a solution $R$ of the Yang-Baxter equation
\be
R_{12}(\lambda)\ R_{13}(\lambda + \lambda')\ R_{23}(\lambda')
= R_{23}(\lambda')\ R_{13}(\lambda + \lambda')\ R_{12}(\lambda) \,.
\label{YB}
\ee 
We consider here the $A_{\n-1}^{(1)}$ $R$ matrix \cite{devega}
\be
R(\lambda) = a \sum_{j} e_{jj}\otimes e_{jj} + b \sum_{j \ne k} 
e_{jj}\otimes e_{kk} + c \sum_{j \ne k} e_{jk}\otimes e_{kj}\ 
e^{-\mu \lambda  \ sign(j-k)} \,,
\label{Rmatrix} 
\ee
where the indices $j \,, k$ take values from 1 to $\n$, and 
$e_{jk}$ are elementary $\n \times \n$ matrices with matrix 
elements $\left( e_{jk} \right)_{ab}=\delta_{ja} \delta_{kb}$. 
Moreover,
\be
a = \sinh \left( \mu  (\lambda + i) \right) \,, \qquad 
b = \sinh ( \mu  \lambda) \,, \qquad 
c = \sinh ( i \mu ) \,,
\label{Rmatrixelements} 
\ee 
where the anisotropy parameter $\mu$ is real.  The $R$ matrix becomes 
$SU(\n)$ invariant for $\mu \rightarrow 0$, and is proportional to the 
permutation matrix for $\lambda=0$.

The corresponding closed-chain transfer matrix $t(\lambda)$ is given 
by
\be
t(\lambda) = \tr_{0} T_{0}(\lambda) \,,
\label{closedtransfer}
\ee
where $\tr_{0}$ denotes trace over the ``auxiliary space'' 0, and
$T_{0}(\lambda)$ is the monodromy matrix
\be
T_{0}(\lambda) = R_{0N}(\lambda) \cdots  R_{01}(\lambda) \,. 
\label{monodromy1}
\ee
We restrict the number of spins, $N$, to be an even number.

The closed-chain Hamiltonian ${\cal H}$ is given by the logarithmic 
derivative of the transfer matrix at $\lambda=0$,
\be
{\cal H}  \sim   {d\over d\lambda } \log t(\lambda) \Big\vert_{\lambda=0}
= {1 \over 2} ( \sum_{n=1}^{N-1} h_{n\,, n+1} + h_{N \,, 1} ) \,,
\ee 
where the two-site Hamiltonian $h$ is given by
\be
h = -\sum_{j \ne k} e_{jj} \otimes e_{kk}\ e^{-i \mu sign(j - k)} + 
\sum_{j\ne k} e_{jk} \otimes e_{kj} \,. 
\label{twosite}
\ee
Here we restrict $\mu$ to lie in the range $\mu \in (0 \,, \pi)$. 
\footnote{The Hamiltonian with $\mu \in (\pi \,, 2\pi)$ is related to 
the parity-transformed Hamiltonian with $\mu \in (0 \,, \pi)$.  
Explicitly, ${\cal H}(\mu) = \Pi {\cal H}(\mu') \Pi$, where 
$\mu' = 2\pi - \mu$, and $\Pi$ is the parity \cite{DN3} operator.  An 
alternative approach (followed, e.g., in \cite{FS},\cite{DN1}) is to restrict 
$\mu$ even further to the range $\mu \in (0 \,, {\pi\over 2})$ and to 
consider both signs of the Hamiltonian.} As for the $A_{\n-1}^{(1)}$ 
Toda field theory with imaginary coupling, the Hamiltonian for $\n > 
2$ is not Hermitian.  Nevertheless, we find in our study of 
low-lying states (see Sec.  3) only real energy eigenvalues.

The transfer matrix has an exact $U(1)^{\n -1}$ symmetry. Indeed, let
\be
s^{(k)} = e_{kk} - e_{k+1 \,, k+1}  \,, \qquad  
k = 1 \,, \ldots \,, \n-1 \,,
\ee
be the Cartan generators in the defining representation of $SU(\n)$; 
and let $s^{(k)}_{n}$ be the corresponding generators at site $n$,
\be
s^{(k)}_{n} = 1 \otimes \ldots \otimes 1 \otimes 
\stackrel{\stackrel{n^{th}}{\downarrow}}{s^{(k)}} \otimes 1 
\otimes \ldots \otimes 1 \,, \qquad n = 1 \,, \ldots \,, N \,.
\ee
We denote by $S^{(k)}$ the corresponding ``total'' generators acting 
on the full space of states
\be
S^{(k)} = \sum_{n=1}^{N} s^{(k)}_{n} \,, \qquad k = 1 \,, \ldots \,, \n-1 \,.
\label{cartan}
\ee 
The transfer matrix has the symmetry
\be
\left[ t(\lambda)  \,, S^{(k)} \right] = 0 \,, 
\qquad k = 1 \,, \ldots \,, \n-1 \,.
\ee 

The eigenstates and eigenvalues of the transfer matrix have been 
determined \cite{sutherland},\cite{kulish/reshetikhin},\cite{devega} 
by the nested algebraic Bethe Ansatz method.  
Indeed, the states are constructed using certain creation operators 
depending on the solutions $\{ \lambda_{\alpha}^{(j)} \}$ of the Bethe 
Ansatz equations
\be
1 &=& \prod_{\beta=1}^{M^{(j-1)}} 
e_{-1}(\lambda_{\alpha}^{(j)} - \lambda_{\beta}^{(j-1)}; \mu)
\prod_{\stackrel{\scriptstyle\beta=1}{\scriptstyle\beta \ne \alpha}}^{M^{(j)}} 
e_{2}(\lambda_{\alpha}^{(j)} - \lambda_{\beta}^{(j)}; \mu)
\prod_{\beta=1}^{M^{(j+1)}} 
e_{-1}(\lambda_{\alpha}^{(j)} - \lambda_{\beta}^{(j+1)}; \mu) \non \\ 
& &  \qquad \qquad 
\alpha = 1 \,, \ldots \,, M^{(j)} \,, \qquad  
j = 1\,, \ldots \,, \n-1  \,,
\label{closedBAE}
\ee
where 
\be
e_n(\lambda; \mu) = 
{\sinh  \mu \left( \lambda + {in\over 2} \right) 
\over           
 \sinh  \mu \left( \lambda - {in\over 2} \right) } \,,
\ee
and $M^{(0)} = N \,, \quad M^{(\n)} = 0 \,, \quad 
\lambda_{\alpha}^{(0)} = \lambda_{\alpha}^{(\n)} = 0 \,.$ 
The corresponding energy, momentum, and $S^{(k)}$ eigenvalues are given by
\be
E &=& - \sin^{2} \mu  \sum_{\alpha=1}^{M^{(1)}} 
{1\over \cosh (2 \mu \lambda_{\alpha}^{(1)}) - \cos \mu }  
\,, \label{energyBA}  \\
P &=& {1 \over i} \sum_{\alpha=1}^{M^{(1)}} 
\log e_{1}(\lambda_{\alpha}^{(1)} \,; \mu) \quad (\mbox{mod } 2 \pi) 
\,, \label{momentumBA}  \\
S^{(k)} &=& M^{(k-1)} + M^{(k+1)} - 2 M^{(k)} \,.
\label{cartanBA}
\ee 

\subsection{Open chain}

In addition to an $R$ matrix, the construction of an integrable open 
spin chain requires also a solution $K$ of the boundary Yang-Baxter 
equation
\cite{cherednik}
\be
R_{12}(\lambda_{1}-\lambda_{2}) K_{1}(\lambda_{1}) R_{21}(\lambda_{1}+\lambda_{2})
K_{2}(\lambda_{2}) =
K_{2}(\lambda_{2}) R_{12}(\lambda_{1}+\lambda_{2}) K_{1}(\lambda_{1}) 
R_{21}(\lambda_{1}-\lambda_{2}) \,.
\label{boundaryYB}
\ee 
We consider here the diagonal $A_{\n-1}^{(1)}$ $K$ matrix \cite{DVGR}
\be
K_{(l)}(\lambda \,, \xi) = \alpha e^{\mu \lambda} \sum_{j=1}^{l} e_{jj} 
+ \beta e^{- \mu \lambda} \sum_{j=l+1}^{\n} e_{jj} \,,
\label{Kmatrix}
\ee
where
\be 
\alpha  = \sinh \left( \mu ( i\xi - \lambda ) \right) \,, 
\qquad \beta = \sinh \left( \mu ( i\xi + \lambda ) \right)   \,,
\label{Kmatrixelements}
\ee
where $\xi$ is a parameter and $l \in \{ 1 \,, \ldots \,, \n-1 \}$. 

The corresponding open-chain transfer matrix 
$t_{(l)}(\lambda \,, \xi_{-} \,, \xi_{+})$ is 
given by \cite{sklyanin}-\cite{DVGR},\cite{DN0}
\be
t_{(l)}(\lambda \,, \xi_{-} \,, \xi_{+}) = 
\tr_{0} M_{0}\ K_{(l)\ 0}(-\lambda-\rho \,, \xi_{+} - {\n\over 2})\  
T_{0}(\lambda)\  K_{(l)\ 0}(\lambda \,, \xi_{-})\ \hat T_{0}(\lambda)\,,
\label{opentransfer}
\ee
where the monodromy matrix $T_{0}(\lambda)$ is given by 
(\ref{monodromy1}), and $\hat T_{0}(\lambda)$ is given by
\be
\hat T_{0}(\lambda) = R_{10}(\lambda) \cdots  R_{N0}(\lambda) \,.
\label{monodromy2}
\ee
Moreover, $M$ is the matrix in the crossing-unitarity relation
\be
R_{12}(-\rho - \lambda)^{t_{1}} M_{1} 
R_{12}(-\rho + \lambda)^{t_{2}} M_{1}^{-1} \propto 1 \,,
\label{bulkcrossunit0}
\ee
and is given by the $\n \times \n$ matrix 
$M_{j k} = \delta_{j k} e^{i \mu (\n - 2 j + 1) }$, and $\rho = i\n/ 2$.
Indeed, it can be shown that this transfer matrix has the 
commutativity property
\be
\left[ t_{(l)}(\lambda \,, \xi_{-} \,, \xi_{+}) \,, 
t_{(l)}(\lambda' \,, \xi_{-} \,, \xi_{+}) \right] = 0 \,,
\label{commutativity}
\ee 
and its derivative at $\lambda=0$ gives the Hamiltonian, 
\be
{\cal H} \sim   {d\over d\lambda } 
t_{(l)}(\lambda \,, \xi_{-} \,, \xi_{+})
\Big\vert_{\lambda=0} 
=  {1\over 2}\sum_{n=1}^{N-1}h_{n\,, n+1} + 
c(\xi_{-}) P_{(l)\ 1} - c(\xi_{+} - \n + l) P_{(l)\ N}  \,,
\ee 
where the two-site Hamiltonian $h$ is given by (\ref{twosite}), and
\be
c(\xi) ={1\over 4}\sin \mu \left(i - \cot(\mu \xi) \right) \,, \qquad
P_{(l)}= \sum_{j=1}^{l} e_{jj} - \sum_{j=l+1}^{\n} e_{jj} \,.
\ee
In particular, for $\n=2$ (and therefore $l=1$), the Hamiltonian is 
given by \cite{ABBBQ},\cite{sklyanin}
\be
{\cal H} =  {1 \over 4 }
\Big \{ \sum_{n=1}^{N-1} \left( 
\sigma^{x}_n \sigma^{x}_{n+1}
+ \sigma^{y}_n \sigma^{y}_{n+1} 
+ \cos \mu  \sigma^{z}_n \sigma^{z}_{n+1} \right)  
- \sin \mu \cot (\mu \xi_-) \sigma^z_1 
+ \sin \mu \cot \left( \mu (\xi_+ - 1) \right) \sigma^z_N 
\Big\} \,. \label{xxz} 
\ee
Evidently, the parameters $\xi_{\pm}$ correspond to boundary fields. 

An important feature \cite{DN0} of this model is its exact quantum 
algebra symmetry $U_{q}\left( SU(l) \right)$ $\times$ $U_{q}\left( 
SU(\n-l) \right)$ $\times$ $U(1)$.  Indeed, let $J^{\pm (k)}$ 
$( k = 1 \,, \ldots \,, \n-1 )$ be raising/lowering 
operators of the quantum algebra $U_{q}\left(SU(\n) \right)$ which 
act on the full space of states, and which obey the 
commutation relations
\be
\left[ J^{+(k)} \,, J^{-(j)} \right] = \delta_{k,j} \left[ S^{(k)} 
\right]_{q} \,, 
\qquad 
\left[ S^{(k)} \,, J^{\pm (j)} \right] = \pm \left( 2 \delta_{k,j} 
-  \delta_{k-1,j} - \delta_{k+1,j} \right) J^{\pm (j)} \,,
\label{algebra}
\ee 
where $\left[ x \right]_{q} \equiv ( q^{x} -  q^{-x})/(q - q^{-1})$,
and $S^{(k)}$ is given by Eq. (\ref{cartan}). For a given value of $l$, 
\be
\left[ t_{(l)}(\lambda \,, \xi_{-} \,, \xi_{+}) \,, S^{(k)} 
\right] &=& 
0 \,, \qquad k = 1 \,, \ldots \,, \n-1 \,, \non   \\
\left[ t_{(l)}(\lambda \,, \xi_{-} \,, \xi_{+}) \,, J^{\pm (k)} 
\right] &=& 
0 \,, \qquad k \ne l \,, 
\label{qsymmetry}
\ee 
where $q=e^{-i\mu}$.  Had we taken the $K$ matrix to be the identity 
matrix $K(\lambda)=1$ instead of (\ref{Kmatrix}), then the 
corresponding transfer matrix would have the full $U_{q}\left(SU(\n) 
\right)$ symmetry \cite{MN}.

The model also has \cite{DN0} a ``duality'' symmetry which relates 
$l \leftrightarrow \n - l$, 
\be
{\cal U}^{l}\ t_{(l)}(\lambda\,, \xi_{-} \,, \xi_{+})\ {\cal U}^{-l} 
\propto  t_{(\n-l)}(\lambda\,, -\xi_{-} \,, -\xi_{+} + \n) \,,
\label{duality}
\ee
where ${\cal U}^{\n} = 1$.   This is a remnant of the 
cyclic ($Z_{\n}$) symmetry \cite{devega}, \cite{belavin} of the 
$A_{\n-1}^{(1)}$ $R$ matrix.

The Bethe Ansatz equations are \cite{DVGR}
\be
1 &=& \left[ e_{2\xi_{-} + l} (\lambda_{\alpha}^{(l)}; \mu)\ 
e_{-\left( 2\xi_{+} - 2 \n + l \right)}(\lambda_{\alpha}^{(l)}; \mu)\  
\delta_{l,j} + \left( 1 - \delta_{l,j} \right) \right]  \non \\  
& & \times \prod_{\beta=1}^{M^{(j-1)}} 
e_{-1}(\lambda_{\alpha}^{(j)} - \lambda_{\beta}^{(j-1)}; \mu)\ 
e_{-1}(\lambda_{\alpha}^{(j)} + \lambda_{\beta}^{(j-1)}; \mu)
\prod_{\stackrel{\scriptstyle\beta=1}{\scriptstyle\beta \ne \alpha}}^{M^{(j)}} 
e_{2}(\lambda_{\alpha}^{(j)} - \lambda_{\beta}^{(j)}; \mu)\
e_{2}(\lambda_{\alpha}^{(j)} + \lambda_{\beta}^{(j)}; \mu) \non \\ 
& & \times \prod_{\beta=1}^{M^{(j+1)}} 
e_{-1}(\lambda_{\alpha}^{(j)} - \lambda_{\beta}^{(j+1)}; \mu)\
e_{-1}(\lambda_{\alpha}^{(j)} + \lambda_{\beta}^{(j+1)}; \mu) \non \\ 
& &  \qquad \qquad 
\alpha = 1 \,, \ldots \,, M^{(j)} \,, \qquad  
j = 1\,, \ldots \,, \n-1  \,.
\label{openBAE}
\ee
The energy is given by (\ref{energyBA}) (plus terms that are 
independent of $\{ \lambda_{\alpha}^{(j)} \}$), and the $S^{(k)}$ 
eigenvalues are again given by Eq.  (\ref{cartanBA}).

\section{Bulk $S$ matrix}

In order to investigate bulk properties, we consider the periodic 
chain with Bethe Ansatz Eqs.  (\ref{closedBAE}).  We assume that the 
ground state is the Bethe Ansatz state with no holes, i.e., with 
$\n-1$ filled real Fermi seas.  Holes in these seas correspond to 
``solitons''.  Indeed, the Bethe Ansatz state with one hole in the 
$j^{th}$ sea is a particle-like excited state which belongs to the 
fundamental representation $[j]$ of $U_{q}\left( SU(\n) \right)$, 
corresponding to a Young tableau with a single column of $j$ boxes.  
Such an excitation with rapidity $\tilde\lambda^{(j)}$ has energy
${\pi \sin \mu \over \mu} s^{(j)}(\tilde\lambda^{(j)})$ and 
momentum $p^{(j)}(\tilde\lambda^{(j)})$, where 
\be
s^{(j)}(\lambda) = {1\over \n}{\sin( {\pi\over \n}(\n - j))\over 
\cos({\pi\over \n}(\n - j)) + \cosh({2 \pi \lambda\over \n})} \,,
\ee 
and the momentum satisfies
\be
{1\over 2\pi} {d\over d\lambda} p^{(j)}(\lambda) = s^{(j)}(\lambda) \,.
\ee 
Note that $s^{(j)}(\lambda)$ has the periodicity $\lambda \rightarrow 
\lambda + i \n$, which suggests that the physical strip is $0 \le  \Im 
m\ \lambda \le  {\n\over 2}$.

Exact bulk scattering matrices can be computed using a generalization 
of the method of Korepin \cite{korepin} and Andrei-Destri \cite{AD}.  
We define the two-particle $S$ matrix $\R^{[j] \otimes [k]}$ for 
particles of type $[j]$ and $[k]$ by the momentum quantization 
condition
\be
\left( e^{ip^{(j)}(\tilde\lambda^{(j)}) N} 
\R^{[j]\otimes[k]} - 1 \right) 
|\tilde\lambda^{(j)}\,, \tilde\lambda^{(k)} \rangle = 0 \,.
\label{quantization}
\ee 
For the scalar factor, we obtain (cf. \cite{DN2})
\be
\R_{0}^{[j] \otimes [k]} &\sim& \exp \left\{ i 2\pi N 
\int_{-\infty}^{\tilde\lambda^{(j)}}
\left( \sigma^{(j)}(\lambda) - s^{(j)}(\lambda) \right) 
d\lambda \right\} \non \\ 
&=& \exp \left\{2\int_{0}^{\infty} {d \omega \over \omega}
\sinh \left( i \omega (\tilde\lambda^{(j)} - \tilde\lambda^{(k)}) \right) 
\left( \delta_{jk} -\hat R_{jk}(\omega) \right) \right\} \,,
\label{bulkS}
\ee 
where $\sigma^{(j)}(\lambda)$ is the density of Bethe Ansatz roots $\{ 
\lambda^{(j)}_{\alpha} \}$ for the state with holes of rapidities 
$\tilde\lambda^{(j)}$ and $\tilde\lambda^{(k)}$ in the $j^{th}$ and 
$k^{th}$ seas, respectively; and (see, e.g., \cite{devega})
\be
\hat R_{j j'}(\omega)=
{\sinh \left({\nu \omega\over 2}\right)\ \sinh \left({j_{<} \omega\over 2}
\right)\ 
\sinh \left( (\n - j_{>}){\omega\over 2} \right) \over
\sinh \left( (\nu-1) {\omega\over 2}\right)\ \sinh \left({\n \omega\over 2}
 \right)\ 
\sinh \left( {\omega\over 2} \right)} \,,
\ee
where $\nu ={\pi\over \mu}$, $j_{>}=\max(j \,, j')$ and $j_{<}=\min(j \,, j')$.
Note that $\mu \in (0 \,, \pi)$ implies $\nu > 1$.

Let us focus on the particular case of two holes of type $[1]$. The 
scalar factor is given by
\be
\R_{0}^{[1] \otimes [1]} = \exp \left\{
2\int_{0}^{\infty} {d \omega \over \omega}
\sinh \left( i \omega \tilde\lambda \right) 
{\sinh \left( {\omega\over 2} \right)\ \sinh\left( (\nu - \n){\omega\over 2} 
\right)
\over
\sinh \left( (\nu-1){\omega\over 2} \right)\ \sinh\left({\n \omega\over 2} 
\right)} \right\} \,,
\label{bulkresult1}
\ee 
where $\tilde\lambda = \tilde\lambda_{1}^{(1)} - 
\tilde\lambda_{2}^{(1)}$.
Moreover, let us consider the following Bethe Ansatz states:
\begin{enumerate}
    
    \item[(a)] two holes and one 2-string (4-string) of positive 
    parity in the first sea if $\nu > {3\over 2}$ ( $\nu < {3\over 
    2}$, respectively)

    \item[(b)] two holes and one 1-string (3-string) of negative 
    parity in the first sea if $\nu > 2$ ( $\nu < 2$, respectively)
    
\end{enumerate}
The corresponding $S$ matrix elements are given by \footnote{Up to 
signs, which are not easy to obtain from the Bethe Ansatz, but which 
can be fixed by requiring that for $\tilde\lambda=0$ the amplitudes be 
equal to $+1$ and $-1$ respectively; i.e., that the matrix $\R(0)$ 
(see below) be equal to the permutation matrix.}
\be
\R_{(a)} = 
{\sin \left( {\pi\over 2(\nu -1)}(1 - i\tilde\lambda) \right)
\over
 \sin \left( {\pi\over 2(\nu -1)}(1 + i\tilde\lambda) \right)}\
\R_{0}^{[1] \otimes [1]} \,, \qquad 
\R_{(b)} = 
-{\cos \left( {\pi\over 2(\nu -1)}(1 - i\tilde\lambda) \right)
\over
 \cos \left( {\pi\over 2(\nu -1)}(1 + i\tilde\lambda) \right)}\ 
\R_{0}^{[1] \otimes [1]} \,.
\ee
It still remains an open problem to determine all $\n^{2}$ 
two-particle states.  However, let us assume that the scattering 
matrix $\R^{[1] \otimes [1]}$ has the same structure as the $R$ matrix 
(\ref{Rmatrix}); i.e.,
\be
\R(\tilde\lambda)^{[1] \otimes [1]} = 
\aaa \sum_{j} e_{jj}\otimes e_{jj} + \bbb \sum_{j \ne k} 
e_{jj}\otimes e_{kk} + \ccc \sum_{j \ne k} e_{jk}\otimes e_{kj}\ 
e^{-x \tilde\lambda  \ sign(j-k)} \,,
\label{bulkSmatrix} 
\ee
where $\aaa \,, \bbb \,, \ccc \,, x$ are to be determined. This matrix has 
the three distinct eigenvalues $\aaa \,, \bbb + \ccc \,, \bbb - \ccc$, which 
we identify with the three amplitudes 
$\R_{0}^{[1] \otimes [1]} \,, \R_{(a)} \,, \R_{(b)}$, respectively. 
Note that the eigenvalues of $\R^{[1] \otimes [1]}$ are independent of $x$. That is, 
the value of $x$ cannot be determined by considering Bethe Ansatz states with 
two holes. We fix the value of $x$ (up to a sign) by requiring that 
$\R^{[1] \otimes [1]}$ obey the Yang-Baxter Eq. (\ref{YB}).\footnote{Presumably, 
one can alternatively consider Bethe Ansatz states with three holes, 
and require factorization of the three-particle $S$ matrix into a 
product of two-particle $S$ matrices, as was done for the XXX model in 
\cite{DMN}.} In this way, we obtain
\be
\aaa = \sin \Big( {\pi\over \nu -1}(1 + i\tilde\lambda) \Big) 
r(\tilde\lambda) \,, \quad  
\bbb = -\sin \Big( {i \pi \tilde\lambda\over \nu -1}\Big) 
r(\tilde\lambda) \,, \quad  
\ccc = \sin \Big( {\pi \over \nu -1}\Big) 
r(\tilde\lambda)  \,, \quad  
x  = {\pi \over \nu -1}  \,,
\label{bulkresult2}
\ee
where
\be
r(\tilde\lambda)={\R_{0}^{[1] \otimes [1]} \over 
\sin \Big( {\pi\over \nu -1}(1 + i\tilde\lambda) \Big)} \,.
\label{bulkresult3}
\ee 
For $\n=2$, this $S$ matrix agrees with the known result for the XXZ 
chain \cite{KR}, \cite{DN3}.  Moreover, the scalar factor 
(\ref{bulkresult1}) coincides with the expression given in Eq.  (3.29) 
of Ref.  \cite{gandenberger1} for the $a_{n}^{(1)}$ scalar factor in 
affine Toda field theory with imaginary coupling, provided we make the 
following identifications:
\be
n\leftrightarrow \n - 1 \,, \qquad \lambda \leftrightarrow 
{1\over \nu -1} \,, \qquad \mu \leftrightarrow {-i \tilde\lambda\over \nu -1} \,.
\ee 

It has been observed \cite{BL}-\cite{gandenberger1} that (after a 
certain gauge transformation) this $S$ matrix has a $U_{q}\left(SU(\n) 
\right)$ symmetry with $q=e^{i \pi\over \nu -1}$.  On the other hand, 
from (\ref{qsymmetry}) we see that the finite-size transfer matrix has 
an ``approximate'' $U_{q}\left(SU(\n) \right)$ symmetry with 
$q=e^{i\pi\over \nu}$.  Evidently, as a result of filling the Fermi 
seas and taking the thermodynamic limit, the value of $q$ becomes 
renormalized. (See also e.g. \cite{DDV1}.)

\section{Boundary $S$ matrix}

We define the boundary $S$ matrices $\K^{\mp}_{(l)\ [j]}$ for a 
particle of type $[j]$ by the quantization condition \cite{GMN}
\be
\left( e^{i 2 p^{(j)}(\tilde\lambda^{(j)}) N} 
\K^{+}_{(l)\ [j]}\ \K^{-}_{(l)\ [j]}- 1 \right) 
|\tilde\lambda^{(j)} \rangle = 0 \,.
\label{quantizationopen}
\ee 
For simplicity, we consider only the case of a hole of type $[1]$.  
The quantum-algebra symmetry (\ref{qsymmetry}) implies that the 
boundary $S$ matrices $\K^{\mp}_{(l)\ [1]}$ are diagonal 
$\n \times \n$ matrices of the same form as the $K$ matrix 
(\ref{Kmatrix}),
\be
\K^{\mp}_{(l)\ [1]} =  
\alpha^{\mp}_{(l)} e^{\pm y \tilde\lambda^{(1)}} \sum_{j=1}^{l} e_{jj} 
+ \beta^{\mp}_{(l)} e^{\mp y \tilde\lambda^{(1)}}\sum_{j=l+1}^{\n} e_{jj} \,,
\label{form}
\ee
where $\alpha^{\mp}_{(l)}$, $\beta^{\mp}_{(l)}$, $y$ are to be 
determined. \footnote{The factors $e^{\pm y \tilde\lambda^{(1)}}$ in 
the boundary $S$ matrices were unfortunately omitted in \cite{DN0}.}
We have the relation
\be
\alpha^{+}_{(l)}\ \alpha^{-}_{(l)}  \sim   
\exp \left\{ i 2\pi N \int_{0}^{\tilde\lambda^{(1)}}
\left( \sigma_{(l)}^{(1)}(\lambda) - 2 s^{(1)}(\lambda) \right) 
d\lambda \right\} \,,
\label{openrel}
\ee 
where $\sigma_{(l)}^{(1)}(\lambda)$ is the density of Bethe Ansatz roots 
$\{ \lambda^{(1)}_{\alpha} \}$ for the state with one hole of 
rapidity $\tilde\lambda^{(1)}$ in the first sea.
From the Bethe Ansatz Eqs.  (\ref{openBAE}), we find (cf.  
\cite{DN0},\cite{DN2})
\be
\alpha^{-}_{(l)} = i \cosh \Big( {\pi\over \nu - 1} ( \tilde \lambda^{(1)}
- {i\over 2}(\nu - 2 \xi_{-}) ) \Big)\ k(\tilde\lambda^{(1)} \,, \xi_{-}) 
\,,
\label{alpha}
\ee 
where
\be 
k(\tilde\lambda \,, \xi) &=& k_{0}(\tilde\lambda )\ 
k_{1}(\tilde\lambda \,, \xi) \,, 
\label{boundscalarfactor1}
\ee 
with
\be
k_{0}(\tilde\lambda ) & = & \exp \Bigg\{
2 \int_{0}^{\infty}{d\omega\over \omega} 
\sinh \left( 2 i \omega \tilde\lambda \right) 
{\sinh \left( (\n +1){\omega\over 2} \right)\
\sinh \left( (\nu - \n){\omega\over 2} \right) \over 
\sinh \left( \n \omega \right)
\sinh \left( (\nu - 1){\omega \over 2} \right)} \Bigg\} \,, \non \\
k_{1}(\tilde\lambda \,, \xi) & = & 
{1\over i \cosh \Big( {\pi\over \nu - 1} ( \tilde\lambda
- {i\over 2}(\nu - 2 \xi) ) \Big)} \non  \\ 
&\times&\exp \Bigg\{
-2 \int_{0}^{\infty}{d\omega\over \omega} 
\sinh \left( 2 i \omega \tilde\lambda \right) 
{\sinh \left( (\n - l) \omega \right)\ 
\sinh \left( (\nu - 2 \xi - l) \omega \right) \over
\sinh\left( (\nu - 1) \omega \right)\ \sinh \left(\n  \omega \right)} 
\Bigg\}
\,. \label{boundscalarfactor2}
\ee 
Moreover, with the help of the ``duality'' symmetry (\ref{duality}), we obtain
\be
\beta^{-}_{(l)} =  i \cosh \Big( {\pi\over \nu - 1} ( \tilde \lambda^{(1)}
+ {i\over 2}(\nu - 2 \xi_{-}) ) \Big)\ k(\tilde\lambda^{(1)} \,, \xi_{-})
\label{beta}
\,.
\ee
Our approach for computing $\alpha^{-}_{(l)}$ and $\beta^{-}_{(l)}$ 
involves performing certain Fourier transformations, which leads to the 
following restriction of parameters: $\n < 2 \xi_{-} + l < 2 \nu$. 
In particular, this requires $\nu > {\n\over 2}$.

The parameter $y$ can be determined using the boundary Yang-Baxter Eq.  
(\ref{boundaryYB}) (in parallel with our analysis of the parameter $x$ 
in the bulk $S$ matrix), and we obtain the value 
\be 
y = {\pi\over \nu-1} \,.
\ee
Our expression for the boundary $S$ matrix agrees with 
\cite{GZ}-\cite{DN1} \footnote{In \cite{DN1} we use a slightly 
different definition of the boundary parameters $\xi_{\mp}$.} 
for $\n=2$, and with \cite{DN2} in the isotropic 
limit $\nu \rightarrow \infty$. A further check on our result is 
performed in the following section. Finally, we remark that for 
$\K^{+}_{(l)\ [1]}$ there are similar expressions involving $\xi_{+}$.

\section{Functional equations for boundary $S$ matrix}

Thus far, we have pursued the direct Bethe-Ansatz calculation of $S$ 
matrices.  Alternatively, $S$ matrices can be found (up to CDD 
ambiguities) by the ``bootstrap'' approach \cite{ZZ}.  Indeed, the 
bulk $S$ matrix scalar factor (\ref{bulkresult1}) was first obtained 
for affine Toda theory in this way \cite{BL}-\cite{gandenberger1}.

A natural question is whether the boundary $S$ matrix can also 
be obtained by a bootstrap approach.  For the case that the bulk $S$ 
matrix has crossing symmetry, Ghoshal and Zamolodchikov \cite{GZ} have 
formulated a boundary crossing relation, which -- together with the 
boundary unitarity relation -- determines the scalar factor of the 
boundary $S$ matrix, up to a boundary CDD ambiguity.  Unfortunately, 
the $A_{\n-1}^{(1)}$ bulk $S$ matrix does not have crossing symmetry 
for $\n >2$, and hence, the Ghoshal-Zamolodchikov relation cannot be 
directly applied.

In this section, we generalize the Ghoshal-Zamolodchikov boundary 
crossing relation to the $A_{\n-1}^{(1)}$ case, and we find that the 
boundary $S$ matrix scalar factor 
(\ref{boundscalarfactor1}),(\ref{boundscalarfactor2}) can indeed 
be obtained by this bootstrap approach.  The basic observation is that, 
although the bulk $S$ matrix lacks crossing symmetry, it does have 
crossing-unitarity symmetry; and one expects that a corresponding property 
should hold for the boundary $S$ matrix. \footnote{A different 
generalization of the Ghoshal-Zamolodchikov boundary crossing relation 
has recently been discussed in Ref.  \cite{HSY}.}

We begin by briefly reviewing the case that the bulk $S$ matrix does 
have crossing symmetry, and rewriting the Ghoshal-Zamolodchikov 
boundary crossing relation in matrix form.  We assume that the 
bulk $S$ matrix $\R(\tilde\lambda)$ has $PT$ symmetry,
\be
\R_{12}(\tilde\lambda)^{t_{1}t_{2}} = \R_{21}(\tilde\lambda) \,,
\label{PT}
\ee
where $\R_{21}={\cal P}_{12} \R_{12} {\cal P}_{12}$, ${\cal P}$ is the 
permutation matrix, and $t_{i}$ denotes transposition in the $i^{th}$ 
space. We write the bulk crossing relation as
\be
\R_{12}(\tilde\lambda)^{t_{2}} = V_{1} \R_{12}(\rho - \tilde\lambda) V_{1} \,,
\label{bulkcrossing}
\ee 
where the crossing matrix $V$ satisfies $V^{2}=1$, and $\rho$ is the 
crossing parameter.   The unitarity relation is
\be
\R_{12}(\tilde\lambda) \R_{21}(-\tilde\lambda) = 1 \,.
\label{bulkunitarity}
\ee
Combining the three relations (\ref{PT})-(\ref{bulkunitarity}), 
one obtains the crossing-unitarity relation
\be
\R_{12}(\rho - \tilde\lambda)^{t_{1}} \M_{1} 
\R_{12}(\rho + \tilde\lambda)^{t_{2}} \M_{1}^{-1} = 1 \,,
\label{bulkcrossunit}
\ee
where $\M = V^{t} V$.  We observe that the Ghoshal-Zamolodchikov 
boundary crossing relation \cite{GZ} for the boundary $S$ matrix 
$\K(\tilde\lambda)$ is consistent with
\be
\tr_{1} \left[ 
\check \R_{21}(- 2 \tilde\lambda) \M_{1} 
\K_{1}(\rho + \tilde\lambda) \right] 
= \left( V_{2} \K_{2}(-\tilde\lambda) V_{2} \right)^{t_{2}} 
\,,
\label{boundcrossing}
\ee
where $\check \R_{12} = {\cal P}_{12} \R_{12}$, and hence $\check 
\R_{21} = \R_{12} {\cal P}_{12}$.  (For aesthetic reasons, we have 
made a shift in the rapidity variable by $\rho/2$, but this is not 
necessary.)

With the help of the boundary unitarity relation 
\be
\K(\tilde\lambda) \K(-\tilde\lambda) = 1 \,,
\label{boundunitarity}
\ee
one can now obtain the relation
\be
\tr_{13}\left[ \check \R_{21}(-2\tilde\lambda) \M_{1} \K_{1}(\rho+\tilde\lambda) 
\check \R_{23}(2\tilde\lambda) \M_{3} \K_{3}(\rho-\tilde\lambda) \right] =1 \,.
\label{boundcrossunit}
\ee
We shall see below that this relation, which is a boundary 
generalization of the bulk crossing-unitarity relation 
(\ref{bulkcrossunit}), is the desired result.

Let us now turn to the $A_{\n-1}^{(1)}$ bulk $S$ matrix 
$\R^{[1] \otimes [1]}$. Although it does not have crossing symmetry, 
it does obey the crossing-unitarity relation (\ref{bulkcrossunit})
with $\M_{j k} = \delta_{j k} e^{-i (\n - 2 j + 1) { \pi\over \nu -1}}$, 
and $\rho = i\n/ 2$ (cf. Eq. (\ref{bulkcrossunit0})). We observe that
this relation together with the unitarity relation 
(\ref{bulkunitarity}) determine the scalar factor of the bulk $S$ 
matrix. Indeed, substituting the form (\ref{bulkSmatrix}), 
(\ref{bulkresult2}) into these relations, we obtain the following 
functional equations for $r(\tilde\lambda)$:
\be
r({i \n\over 2} + \tilde\lambda)\ 
r({i \n\over 2} - \tilde\lambda) &=&
{1\over \sin \Big( {\pi\over \nu -1}({\n\over 2} + i\tilde\lambda) \Big)
\sin \Big( {\pi\over \nu -1}({\n\over 2} - i\tilde\lambda) \Big)} \,,
\non \\ 
r(\tilde\lambda)\ r( - \tilde\lambda) &=&
{1\over \sin \Big( {\pi\over \nu -1}(1 + i\tilde\lambda) \Big)
\sin \Big( {\pi\over \nu -1}(1 - i\tilde\lambda) \Big)} \,.
\ee
Solving these equations yields the result (\ref{bulkresult3}), 
(\ref{bulkresult1}), up to a CDD factor.

Finally, we turn to the $A_{\n-1}^{(1)}$ boundary $S$ matrix 
$\K^{-}_{(l)\ [1]}$.  Substituting the form (\ref{form}), 
(\ref{alpha}), (\ref{beta}) into the boundary crossing-unitarity 
relation (\ref{boundcrossunit}), we obtain a functional equation for 
the scalar factor $k(\tilde\lambda \,, \xi)$:
\be 
k({i \n\over 2} + \tilde\lambda\,, \xi)\ 
k({i \n\over 2} - \tilde\lambda\,, \xi) =
{\sin \Big( {\pi\over \nu -1}(1 + 2i\tilde\lambda) \Big)
 \sin \Big( {\pi\over \nu -1}(1 - 2i\tilde\lambda) \Big) \over 
\sin \Big( {\pi\over \nu -1}(\n + 2i\tilde\lambda) \Big)
\sin \Big( {\pi\over \nu -1}(\n - 2i\tilde\lambda) \Big)} 
\non \\ 
\times {(-1) \over \cosh \Big( {\pi\over \nu - 1} 
( \tilde \lambda - {i\over 2}(\nu - 2 \xi + \n - 2l) ) \Big)
\cosh \Big( {\pi\over \nu - 1} 
( \tilde \lambda + {i\over 2}(\nu - 2 \xi + \n - 2l) ) \Big)}  
\ee
(up to a rapidity-independent phase, which can be absorbed by a 
redefinition of $\M$).  Similarly, the boundary unitarity relation 
(\ref{boundunitarity}) implies
\be 
k(\tilde\lambda\,, \xi)\ 
k( - \tilde\lambda\,, \xi) =
-{1\over 
\cosh \Big( {\pi\over \nu - 1} 
( \tilde \lambda - {i\over 2}(\nu - 2 \xi) ) \Big)
\cosh \Big( {\pi\over \nu - 1} 
( \tilde \lambda + {i\over 2}(\nu - 2 \xi) ) \Big)} \,.
\ee
Solving these equations yields the result (\ref{boundscalarfactor1}), 
(\ref{boundscalarfactor2}) up to a CDD factor.

\section{Discussion}

The critical $A_{\n-1}^{(1)}$ quantum spin chain shares a number of 
features with $A_{\n-1}^{(1)}$ Toda quantum field theory with 
imaginary coupling.  In particular, the two models have solitonic 
excitations with the same bulk $S$ matrix.  This suggests that the 
spin chain may be a type of discretization (see, e.g., 
\cite{DDV1},\cite{DDV2}) of the affine Toda theory.  This further 
suggests that it may be possible to find ``soliton preserving'' 
integrable boundary conditions for the affine Toda theory which lead 
to the boundary $S$ matrix which we have given here.  Searches for 
such integrable boundary conditions have not yet been successful (see, 
e.g., \cite{corrigan},\cite{gandenberger2}), but progress in this 
direction has recently been reported \cite{delius}.  (Boundary $S$ 
matrices with a structure similar to (\ref{form}) have recently been 
proposed for the $O(N)$ Gross-Neveu and nonlinear sigma models 
\cite{DMM} and for the $SO(N)$ principal chiral model \cite{mackay}.) 
As in the affine Toda theory, the spin chain has various bound states, 
on which we hope to report soon.

\section*{Acknowledgments}

We thank V.  Brazhnikov for helpful discussions.  This work was 
supported in part by the National Science Foundation under Grant 
PHY-9870101.

Note Added: After completion of this work, we became aware of Ref.  
\cite{zinn-justin}, which puts forward the more general conjecture 
that the light-cone continuum limit of the critical $\hat g$ spin 
chain is the $\hat g$ Toda theory with imaginary coupling, for any 
untwisted affine simply-laced Lie algebra $\hat g$.  The result 
(\ref{bulkresult1}) for the scalar factor of the bulk $S$ matrix for 
the $A_{\n-1}^{(1)}$ chain is also obtained there.  We are grateful to 
G.  Takacs for bringing this reference to our attention.

\end{document}